\documentclass{article}
\usepackage{spconf,amsmath,graphicx}
\usepackage{amsmath,amssymb}
\usepackage{multirow}
\usepackage{graphicx}
\usepackage{epstopdf}
\usepackage{hyperref}
\usepackage{xcolor}
\usepackage{soul}
\usepackage{color,soul}
\usepackage{tabularx}
\usepackage{cite}
\usepackage{subcaption}
\usepackage[english]{babel}
\usepackage{enumitem}
\usepackage{breqn}
\usepackage{xcolor}
\usepackage{makecell}
\usepackage{eqparbox}
\usepackage{booktabs}
\usepackage{multirow}
\DeclareMathOperator{\E}{\mathbb{E}}
\newdimen{\algindent}
\usepackage{eso-pic}
\usepackage[linesnumbered,ruled,vlined]{algorithm2e}

\usepackage{bm}

\SetCommentSty{mycommfont}

\SetKwInput{KwInput}{Input}                
\SetKwInput{KwOutput}{Output}              

\title{Adversarial defense for deep speaker recognition using hybrid adversarial training}
\name{\begin{tabular}{c}Monisankha Pal$^{1}$, Arindam Jati$^{1}$, Raghuveer Peri$^{1}$, Chin-Cheng Hsu$^{1}$, Wael AbdAlmageed$^{2}$, \\ Shrikanth Narayanan$^{1,2}$\end{tabular} 
}
\address{
$^{1}$Signal Analysis and Interpretation Laboratory, University of Southern California (USC), USA\\
$^{2}$USC Information Sciences Institute, Marina del Rey, USA
}

\begin{document}
\ninept
\maketitle
\begin{abstract}
Deep neural network based speaker recognition systems can easily be deceived by an adversary using minuscule imperceptible perturbations to the input speech samples. These adversarial attacks pose serious security threats to the speaker recognition systems that use speech biometric. To address this concern, in this work, we propose a new defense mechanism based on a hybrid adversarial training (HAT) setup. In contrast to existing works on countermeasures against adversarial attacks in deep speaker recognition that only use class-boundary information by supervised cross-entropy (CE) loss, we propose to exploit additional information from supervised and unsupervised cues to craft diverse and stronger perturbations for adversarial training. Specifically, we employ multi-task objectives using CE, feature-scattering (FS), and 
margin losses to create adversarial perturbations and include them for adversarial training to enhance the robustness of the model. 
We conduct speaker recognition experiments on the Librispeech dataset, and compare the performance with state-of-the-art projected gradient descent (PGD)-based adversarial training which employs only CE objective.
The proposed HAT improves adversarial accuracy by absolute 3.29\% and 3.18\% for PGD and Carlini-Wagner (CW) attacks respectively, while retaining high accuracy on benign examples.  
\end{abstract}
%
%
\section{Introduction}
\label{sec:intro}

Although deep learning approaches have revolutionized many areas of research, they remain vulnerable to \textit{adversarial attacks} \cite{szegedy2013intriguing, yuan2019adversarial}. These malicious attacks which contain perturbations imperceptible to humans are intentionally crafted by an adversary to deceive a trained deep neural network (DNN). Such adversarial examples that are perceptually indistinguishable from the original examples pose serious security threats to DNN-based speech biometric systems like speaker recognition by drastically degrading its accuracy. Therefore, to mitigate adversarial attacks, designing robust speaker recognition systems with efficient defense mechanisms is an imminent need.
\par
Initial works on adversarial examples to attack a deep learning model have mainly focused on image classification \cite{yuan2019adversarial, hao2020adversarial}. Adversarial attacks are ubiquitous beyond image classification, 
ranging from automatic speech recognition (ASR) \cite{carlini2018audio} 
to speaker recognition \cite{chen2019real}. 
In \cite{carlini2018audio}, 
targeted audio adversarial examples on an end-to-end ASR system were applied to achieve a 100\% attack success rate. Attacks based on psycho-acoustic hiding \cite{schonherr2018adversarial} and auditory masking \cite{qin2019imperceptible} which can produce human imperceptible audio were introduced to deep ASR models with high success rate. 
On the other hand, 
the vulnerability of speaker verification (SV) systems 
to fast gradient sign method (FGSM) attacks 
have been explored in literature 
on methods based on end-to-end DNNs \cite{kreuk2018fooling} and 
i-vectors \cite{li2020adversarial}.
In \cite{qin2019imperceptible}, the authors have employed natural evolution strategy to craft adversarial samples to successfully attack GMM-UBM and i-vector based speaker recognition systems. Recently, a generative network was proposed in \cite{li2020universal}, which learns to yield universal adversarial perturbation to attack SincNet-based speaker recognition system. However, most of the aforementioned works have not used stronger available attack algorithms.  
\par
There are very few defense methods that have been proposed to mitigate adversarial attacks in speaker recognition systems. In \cite{wang2019adversarial}, adversarial regularization based on adversarial examples using FGSM and local distributional smoothness was proposed to enhance the robustness of the end-to-end SV system. A passive defense using spatial smoothing, and a proactive defense using \textit{adversarial training (AT)} were introduced to handle adversarial attacks in SV spoofing countermeasure system \cite{wu2020defense}. In our previous work, we presented an expository study to show the effect of multiple state-of-the-art adversarial attacks on a deep speaker recognition system \cite{jati2020adversarial}. Furthermore, multiple defense techniques resilient against adversarial attacks were also introduced. Although these works demonstrate initial efforts on adversarial defense in deep speaker recognition systems, their efficacy against large set of diverse adversarial attacks have not been explored.
\par
To address these issues, in this work, we first propose a new defense approach through the lens of adversarial training using adversarial perturbations generated by \textit{feature scattering (FS)}. The feature scattering based adversarial training which is unsupervised and avoids the label-leaking \cite{kurakin2016adversarial} issue of supervised schemes was recently introduced in computer vision \cite{zhang2019defense}. First, we adopt and study the effectiveness of the FS-based defense method against adversarial attacks in the speaker recognition context. Second, we improve the adversarial training further by exploiting additional information from the feature space to craft diverse and stronger adversaries. We propose an attack using a \textit{hybrid setup} based on incorporating \textit{multi-task objectives}$:$ class-boundary information (cross-entropy (CE) loss), feature matching distance between original and perturbed samples (FS loss), and the difference between the true class and the most-confusing class posteriors (margin loss) \cite{carlini2017towards}. In this manner, the perturbation can find the blind-spots of the model more effectively and in turn, that is used to learn a robust model through adversarial training. The contributions of this work are (a) a new hybrid adversarial training scheme based on multi-loss objectives, (b) analysis and comparison of results against various state-of-the-art white-box attacks \cite{carlini2019evaluating} (Sec. \ref{sec:white-box attacks}), and (c) experiments on transfer/black-box attacks \cite{carlini2019evaluating}. 

\vspace{-10pt}
\section{Adversarial attack and training in deep speaker recognition}
\label{sec2}
\vspace{-5pt}

Speaker recognition, the task of recognizing the speakers from their spoken utterances, can be performed either for identification or verification \cite{kinnunen2010overview, pal2015robustness}. For closed-set speaker identification, based on training utterances from a set of unique speakers, a DNN model is trained using the classification loss such as cross-entropy (CE), and during testing the input utterance is classified as belonging to one of the speakers seen in training. The task of speaker verification (SV) is to accept or reject an identity claim from a speech sample \cite{jati2019multi}. Although it is an open-set problem, many recent state-of-the-art SV systems are formulated during model training with a classification objective e.g., with a CE loss. 
Let $\mathbf{x} \in \mathbb{R}^d$ denote an original input speech sample with speaker label $y$, the standard speaker recognition model $\boldsymbol{\theta}$ is trained to minimize the CE loss as
\begin{equation}
    \underset{\boldsymbol{\theta}}{\textnormal{argmin}} \hspace{2pt} \E_{(\mathbf{x}, y) \sim S} \mathcal{L}(\mathbf{x}, y; \boldsymbol{\theta}) 
    \vspace{-5pt}
\end{equation}
where $\mathcal{L}(\cdot)$ is the cross-entropy loss, $S$ represents the training set.
\par
The adversarial perturbation $\boldsymbol{\xi}$ is added to $\mathbf{x}$ to generate adversarial sample $\mathbf{x}_{\textnormal{adv}} = \mathbf{x} + \boldsymbol{\xi}$ such that $\Vert \boldsymbol{\xi} \Vert_p < \epsilon$, where $\epsilon$ is the allowed perturbation budget. The perturbation $\boldsymbol{\xi}$ is optimized by maximizing the loss function and this can be solved approximately either using a one-step FGSM \cite{goodfellow2014explaining}, or a multi-step projected gradient descent (PGD) \cite{madry2017towards} approach. The general form of adversary generation can be written as 
\begin{equation}
    \mathbf{x}_{\textnormal{adv}}^{t+1} = \mathcal{P}_{S_{\mathbf{x}}}\left(\mathbf{x}_{\textnormal{adv}}^{t} + \alpha \cdot \textnormal{sign} \left(\nabla_{\mathbf{x}_{\textnormal{adv}}^{t}}\mathcal{L}\left(\mathbf{x}_{\textnormal{adv}}^{t}, y; \boldsymbol{\theta}\right)\right)\right) \label{eq2}
    \vspace{-5pt}
\end{equation} 
where $\mathcal{P}_{S_{\mathbf{x}}}$ is a projection operator as in a standard PGD optimization algorithm, $S_{\mathbf{x}}$ is the set of allowed perturbations, i.e., $l_{\infty}$ or $l_p$ ball around $\mathbf{x}$, $\alpha$ is the step size for gradient descent update. For FGSM attack, the total number of iterations is 1 and for PGD-$T$ attack it is $T$. 
\par
The basic idea of adversarial training is to inject adversarial samples continually generated from each minibatch and perform training in an iterative online fashion using those samples. Adversarial training for defense improves the model robustness by solving a minimax problem as
\begin{equation}
    \underset{\boldsymbol{\theta}}{\textnormal{min}} \left[\max_{\mathbf{x}_{\textnormal{adv}} \in S_{\mathbf{x}}} \mathcal{L}\left(\mathbf{x}_{\textnormal{adv}}, y; \boldsymbol{\theta}\right)\right] 
    \label{eq3}
\end{equation}
where adversarial attack is generated by inner maximization and adversarial training is based on outer minimization of the loss induced by the attack.

\vspace{-10pt}
\section{Feature scattering adversarial training}
\vspace{-5pt}
\label{fea_sca}

Supervised methods using cross-entropy loss for adversary generation move the data points towards the decision boundary and thus disregard the original data manifold structure which might hinder the classification performance. Moreover, supervised methods suffer from label-leaking (adversarial perturbation is highly correlated with the original speaker label and the classifier can directly infer the speaker label from the perturbation without relying on the real content of the data) issue \cite{kurakin2016adversarial}. To address these issues, the feature scattering approach was introduced to generate adversarial samples considering the inter-sample relationship \cite{zhang2019defense}. This unsupervised approach employs optimal transport (OT) \cite{cuturi2013sinkhorn} based feature matching distance between two discrete distributions $\boldsymbol{\mu}, \boldsymbol{\nu}$ 
as the loss function for perturbation generation. The two discrete distributions $\boldsymbol{\mu}$ and $\boldsymbol{\nu}$ are: $\boldsymbol{\mu} = \sum_{i=1}^n \mu_i \delta_{x_i}$ and $\boldsymbol{\nu} = \sum_{i=1}^n \nu_i \delta_{x_{\textnormal{adv}, i}}$, where $\delta_{x}$ is the Dirac function, and $\boldsymbol{\mu}, \boldsymbol{\nu}$ are $n$-dimensional simplex, and $n$ is the minibatch size. The OT distance is the minimum cost of transporting $\boldsymbol{\mu}$ to $\boldsymbol{\nu}$ and can be defined as 
\begin{equation}
    \boldsymbol{\mathcal{D}}(\boldsymbol{\mu}, \boldsymbol{\nu}) = \underset{\textbf{T} \in \Pi(\boldsymbol{\mu}, \boldsymbol{\nu})}{\textnormal{min}} \langle \mathbf{T}, \mathbf{C} \rangle \label{eq4}
\end{equation} 
where $\langle \cdot, \cdot \rangle$ stands for Frobenius dot product, $\textbf{T}$ is the transport matrix and $\mathbf{C}$ denotes cost matrix. The set $\Pi(\boldsymbol{\mu}, \boldsymbol{\nu})$ contains all possible joint probabilities with marginals $\boldsymbol{\mu}(\mathbf{x})$, $\boldsymbol{\nu}(\mathbf{x}_{\textnormal{adv}})$, and $\Pi(\boldsymbol{\mu}, \boldsymbol{\nu}) = \{\mathbf{T} \in \mathcal{R}_{+}^{n \times n}|\mathbf{T}\mathbf{1}_{n} = \boldsymbol{\mu}, \mathbf{T}^{\top}\mathbf{1}_{n} = \boldsymbol{\nu}\}$ \cite{zhang2019defense}. The cost matrix $\mathbf{C}$ is computed based on cosine distance between original and perturbed samples in the feature space, which is defined as
\begin{equation}
    \mathbf{C}_{ij} = 
    1 - \frac{f_{\boldsymbol{\theta}}(\mathbf{x}_i)^{\top} f_{\boldsymbol{\theta}}(\mathbf{x}_{\textnormal{adv}, j})}{\Vert f_{\boldsymbol{\theta}}(\mathbf{x}_i) \Vert_2 \Vert f_{\boldsymbol{\theta}}(\mathbf{x}_{\textnormal{adv}, j})\Vert_2}
\end{equation}
Feature matching distance, as defined in Eq. (\ref{eq4}), is used in the inner maximization loop in Eq. (\ref{eq3}). The adversarial samples for training are thus generated in an iterative fashion as in Eq. (\ref{eq2}).
The adversarial training or the outer minimization procedure of Eq. (\ref{eq3}) uses cross-entropy loss to update the parameters of the model. 

\vspace{-10pt}
\section{Proposed methodology}
\label{sec:majhead}
\vspace{-5pt}

\subsection{Motivation}
\label{ssec:subhead}
\vspace{-5pt}

Optimizing inner maximum of Eq. (\ref{eq3}) based on hand-crafted first-order gradients often leads to local maxima that approximates the optimal adversary. Success of the defense algorithms depends on the quality of the local maximum. Stronger adversarial perturbation may facilitate computing better local maximum \cite{jang2019adversarial}. Moreover, the recognition system can be made more robust by training it with diverse adversarial attacks that can reveal the various blind spots or vulnerabilities of the system. Therefore, in the pursuit of obtaining strong and diverse adversaries, multi-task objective based on multiple supervised/unsupervised cues could be useful, which might also help in regularizing the network. Intuitively, the model can become more robust if it is trained with diverse and stronger attacks \cite{jang2019adversarial}.
 
\vspace{-10pt}
\subsection{Hybrid adversary generation}
\label{sssec:subsubhead}
\vspace{-5pt}
In this work, we propose to jointly utilize information from cross-entropy (CE), feature scattering (FS) and margin losses to craft adversarial perturbations. Specifically, we extract class boundary information, inter-sample relationship by keeping the original data manifold, and information to yield stronger adversaries by adding minimum perturbation. The loss function that is maximized in the proposed hybrid adversarial training (HAT) setup is as follows:
\begin{equation}
    \mathcal{L}_{\textnormal{adv}} = \beta \cdot \mathcal{L}_{\textnormal{CE}} (\mathbf{x}_{\textnormal{adv}}, y) + \gamma \cdot \mathcal{L}_{\textnormal{FS}} (\mathbf{x}, \mathbf{x}_{\textnormal{adv}}) + \zeta \cdot \mathcal{L}_{M} (\mathbf{x}_{\textnormal{adv}}, y) \label{eq6}
\end{equation}
where $\beta$, $\gamma$, $\zeta$ are the tunable hyper-parameters. Here, all the three losses are calculated from the model output or logit space. We describe the individual losses below.

\vspace{-10pt}
\subsubsection{Cross-entropy loss}
\label{sec:print}

The cross-entropy loss pushes the adversarial samples towards the decision boundary \cite{mustafa2020deeply}. 
The objective of the softmax cross-entropy loss is to maximize the dot product between an input feature and its true class representative and can be written as
\begin{equation}
    \mathcal{L}_{\textnormal{CE}} (\mathbf{x}_{\textnormal{adv}}, y) = - \frac{1}{n} \sum_{i=1}^n  \mathbf{1}_{y,i}^{\top} \hspace{2pt} \textnormal{log} [p(f_{\boldsymbol{\theta}}(\mathbf{x}_{\textnormal{adv}, i}))]
\end{equation}
where for $(\mathbf{x}_{\textnormal{adv}}, y)$ pair $\mathbf{1}_y$ is the one-hot encoding of $y$, 
the second term is the log of predicted probability, and $n$ is the minibatch size. 
Maximizing this loss in multiple iterations like PGD-attack \cite{madry2017towards} can yield diverse attacks.

\vspace{-12pt}
\subsubsection{Feature-scattering loss}
\vspace{-5pt}

Since feature-scattering is performed on a batch of samples by leveraging the inter-sample structure, it induces a coupled regularization useful for adversarial training. Nevertheless, it promotes data diversity without altering the data manifold structure much as compared to supervised approach with label-leaking phenomenon. The maximization of feature scattering loss for adversary generation is $\mathcal{L}_{\textnormal{FS}} (\mathbf{x}, \mathbf{x}_{\textnormal{adv}}) = \boldsymbol{\mathcal{D}}(\boldsymbol{\mu}, \boldsymbol{\nu})$ given in Eq. (\ref{eq4}). 

\vspace{-10pt}
\subsubsection{Margin loss}
\vspace{-5pt} 

The margin loss based on Carlini-Wagner (CW) attack tries to find the minimally-distorted perturbation \cite{carlini2017towards, jati2020adversarial}. The objective of this loss is to obtain stronger adversarial attack by minimizing the difference between the logits of the true class of the adversarial sample and the most confusing class that is not the true class, with a margin parameter that controls the confidence of the attack. The margin loss that is maximized in our proposed defense is computed as
\begin{equation}
    \mathcal{L}_{M} (\mathbf{x}_{\textnormal{adv}}, y) = \sum_{i=1}^n - \left[f_{\boldsymbol{\theta}}(\mathbf{x}_{\textnormal{adv}, i})_t - \underset{j \neq t}{\textnormal{max}}(f_{\boldsymbol{\theta}}(\mathbf{x}_{\textnormal{adv}, i})_j) + M \right]_{+}
\end{equation}
where $t$ represents the output node corresponding to the true class $y$, $M$ is the confidence margin, and $[\cdot]_{+}$ represents $\textnormal{max}(\cdot, 0)$ function. 

\vspace{-10pt}
\subsection{Hybrid adversarial training (HAT)}
\vspace{-5pt}
The model training part (outer loop in Algorithm \ref{algo1}) of HAT employs CE losses computed between clean samples and labels, and adversarial samples (using proposed hybrid adversary) and labels. 
The proposed training framework is listed in Algorithm \ref{algo1}.

\begin{algorithm}[!t]
\DontPrintSemicolon
  
  \KwInput{Classifier model $\boldsymbol{\theta}$, Dataset $S$, number of training epochs $N_e$, learning rate $\eta$, attack iterations $T$, $\epsilon = 0.002$, step size $\alpha = \epsilon/5$, $w_1$, $w_2$}
  \KwOutput{model $\boldsymbol{\theta}$}
 
  \For{$k = 1$ to $N_e$}
    {
        \For{\textnormal{random batch} $(\mathbf{x}_i, y_i)_{i=1}^n \sim S$}    
        {
            Initialize $\boldsymbol{\mu} = \sum_{i=1}^n \mu_i \delta_{x_i}$, $\boldsymbol{\nu} = \sum_{i=1}^n \nu_i \delta_{x_{\textnormal{adv}, i}}$, $\Vert \mathbf{x}_{\textnormal{adv}} - \mathbf{x} \Vert_{\infty} < \epsilon$ \\
            \For{$t = 1$ to $T$}  
            {
                Forward pass $\mathbf{x}$ and compute $\mathcal{L}_{\textnormal{adv}}$ \tcp*{Eq.(\ref{eq6})}  
                
                $\mathbf{x}_{\textnormal{adv}}^{t+1} = \mathcal{P}_{S_{\mathbf{x}}}\left(\mathbf{x}_{\textnormal{adv}}^{t} + \alpha \cdot \textnormal{sign} \left(\nabla_{\mathbf{x}_{\textnormal{adv}}^{t}}\mathcal{L}_{\textnormal{adv}}\right)\right)$
            }
            \tcc{Backward pass and update the model parameters to minimize the loss}
            $\mathcal{L}(\mathbf{x}_{\textnormal{adv}}, y; \boldsymbol{\theta}) = w_1 \cdot \mathcal{L}_{\textnormal{CE}} (\mathbf{x}, y) + w_2 \cdot \mathcal{L}_{\textnormal{CE}} (\mathbf{x}_{\textnormal{adv}}, y)$
            
            $\boldsymbol{\theta} \longleftarrow \boldsymbol{\theta} - \eta \cdot \frac{1}{n} \sum_{i=1}^n \nabla_{\boldsymbol{\theta}} \mathcal{L}(\mathbf{x}_{\textnormal{adv}, i}, y_i; \boldsymbol{\theta})$
           
        }
    }
\vspace{-5pt}
\caption{Hybrid adversarial training (HAT)}
\label{algo1}
\end{algorithm}

\vspace{-10pt}
\section{Experimental setup}
\label{sec:page}

\vspace{-5pt}
\subsection{Dataset}
\vspace{-5pt}

We conduct all the speaker recognition experiments on Librispeech \cite{panayotov2015librispeech} dataset. We employ the ``train-clean-100" subset which contains 100 hours of read English speech from 251 unique speakers. For every speaker, we keep 90\% of the utterances for training the recognizer and use the remaining 10\% for inference. This train-test split is kept fixed for all the experiments in this paper.

\vspace{-10pt}
\subsection{Implementation details}
\vspace{-5pt}

The model we choose for our closed-set speaker classification task is based on 1D CNN, which has 8 stacks of convolution layers \cite{jati2020adversarial}. We apply ReLU non-linearity and batch-normalization after every conv layer and max pooling after every alternate layer. The final layer is a linear fully connected layer with 251 (\# speakers) output nodes. It is to be noted that our model also contains a non-trainable digital signal processing front-end as a pre-processor, which is differentiable and extracts log Mel-spectrogram from the raw speech signal. Therefore, our adversarial attacks are applied directly in the time-domain.
\par
We employ SGD with momentum optimizer \cite{ruder2016overview} for training the model with initial learning rate 0.1 until 60 epochs, then reduces it to 0.01 until 90 epochs, and further reduces it to 0.001 through the total 200 epochs. The perturbation budget $\epsilon$ = 0.002 is used in training and testing similar to our previous work \cite{jati2020adversarial}, and it produces adversarial examples with signal-to-noise ratio (SNR) $\sim$ 30dB. During training, for FS loss computation we use Sinkhorn Algorithm \cite{cuturi2013sinkhorn} with regularization of 0.01 by following the original paper \cite{zhang2019defense}. For margin loss, the confidence margin is fixed at 50. Attack iteration $T = 10$ is applied for adversary generation in our HAT. In this work, we set the value of $\beta, \gamma, \zeta, w_1$ and $w_2$ as 1.
\par
For testing, model robustness is evaluated by measuring the accuracy under different adversarial attacks such as white-box FGSM \cite{goodfellow2014explaining}, PGD \cite{madry2017towards}, CW \cite{carlini2017towards} (within the same PGD framework), FS \cite{zhang2019defense} and also few variants of black-box attacks. The attack strength $\epsilon$ here is also 0.002; however, we have shown performance with other $\epsilon$-values as well. We compare proposed HAT10 defense method against other defenses such as FGSM-AT, PGD10-AT, FS10-AT \footnote{Code will be released with the paper.}.

\begin{table*}[!t]
\caption{Accuracy comparison of the proposed defense with other baseline defenses under untargeted, white-box attacks with $\epsilon = 0.002$}
\vspace{-10pt}
\centering
\setlength{\tabcolsep}{8pt}
\begin{tabular}{c|c|c|ccc|ccc|ccc}
\Xhline{2.5\arrayrulewidth}
\multirow{2}{*}{Defense} & \multirow{2}{*}{Clean} & \multirow{2}{*}{FGSM-attack} & \multicolumn{3}{c|}{PGD-attack}                          & \multicolumn{3}{c|}{CW-attack}                           & \multicolumn{3}{c}{FS-attack}  \\
          \cline{4-6} \cline{7-9}  \cline{10-12}              &                        &                       & 10             & 20             & 40             & 10             & 20             & 40             & 10     & 20    & 40                            \\
\Xhline{2.5\arrayrulewidth}
Standard                 & \textbf{99.55}                  & 6.03                  & 0.00           & 0.00           & 0.00           & 0.00           & 0.00           & 0.00           & 0.00   & 0.00  & 0.00                          \\
FGSM-AT                     & 97.64                  & 57.04                 & 14.86          & 11.05          & 9.59           & 17.01          & 14.13          & 12.37          & 81.19  & 79.13 & 77.5                             \\
PGD10-AT                    & 97.30                  & 88.78                 & 78.22          & 76.62          & 75.55          & 77.42          & 76.00          & 75.34          & 96.57  & 96.57 & 96.35                            \\
FS10-AT                     & 96.12                  & 82.85                 & 60.69          & 55.28          & 53.19          & 38.45          & 35.68          & 32.54          & 96.81  & 96.51 & 96.47                            \\
HAT10                    & 97.68                  & \textbf{90.06}                 & \textbf{81.12} & \textbf{79.60} & \textbf{78.84} & \textbf{80.12} & \textbf{79.18} & \textbf{78.52} & \textbf{96.95}  & \textbf{96.90} & \textbf{96.67} \\
\Xhline{2.5\arrayrulewidth}
\end{tabular}\label{table1}
\vspace{-17pt}
\end{table*}

\vspace{-10pt}
\section{Results and discussions}
\label{sec:illust}


\vspace{-5pt}
\subsection{Performance under untargeted white-box attacks}
\label{sec:white-box attacks}

\vspace{-5pt}
\subsubsection{Main results}
\vspace{-5pt}
We compare in Table~\ref{table1} the performance of
the defenseless model (denoted as Standard),
popular AT models,
and our proposed HAT model
under the untargeted white-box attack scenario
where the adversary possesses complete knowledge of the model (e.g., architecture and values of the parameters).
For a fair comparison, we follow the same threat model, i.e., $l_{\infty}$-attacks with $\epsilon = 0.002$ for both training and testing. It is observed that although benign (clean) accuracy of the standard model is very high (99.55\%), it fails drastically under all the white-box attacks. Comparing different defense methods we observe that FGSM-AT is the weakest defense approach. We note that PGD10-AT achieves consistently higher robustness against all the attacks. On the other hand, although FS10-AT approach provides high accuracy for clean, FGSM and FS attacks, its performance degrades under stronger PGD and CW attacks. This could be attributed to the fact that FS based on OT distance increases the loss towards some unknown class and there is no direct mechanism involved to increase inter-class margin to induce stronger attacks. Finally, using our proposed HAT, we obtain significant performance boost over FS method after incorporating multi-task objectives. Furthermore, our defense attains 3.29\% and 3.18\% absolute improvement over PGD10-AT under PGD40 and CW40 adversarial attacks. In addition, HAT yields superior performance over all the defense methods under both seen (PGD10, CW10, FS10) and strong unseen (PGD40, CW40, FS40) attacks. 

\begin{figure}[!t]
 \centering
  \includegraphics[width=0.39\textwidth]{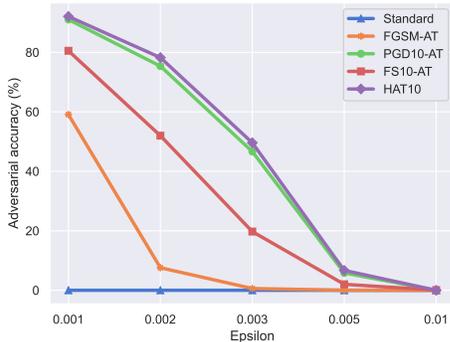}
  \vspace*{-0.42cm}
  \caption{Performance of different defense methods under PGD100 attack with different attack budgets.}
  \label{fig2}
  \vspace{-15pt}
\end{figure}

\begin{figure}[!t]
 \centering
  \includegraphics[width=0.39\textwidth]{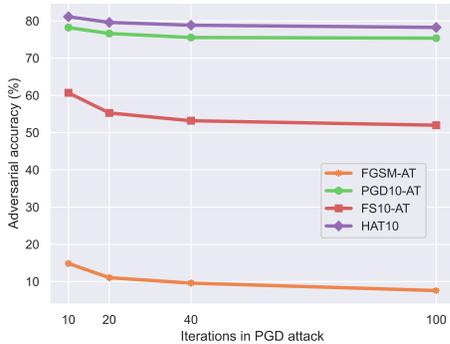}
  \vspace*{-0.42cm}
  \caption{Performance of different defense methods under PGD attack with different attack iterations at fix $\epsilon = 0.002$.}
  \label{fig3}
  \vspace{-15pt}
\end{figure}

\vspace{-12pt}
\subsubsection{Analysis}
\vspace{-5pt}

Fig.~\ref{fig2} shows the performance variation of different defense methods with respect to the change in perturbation ($\epsilon$) strength. We vary the epsilon value from 0.001 to 0.01 to simulate the PGD100 attack while testing, however, all the trained models are based on same $\epsilon$ = 0.002. We observe that the trend of the curves are downward with an increase in perturbation budget. Among the different defenses, we see that the proposed HAT10 defense is slightly superior to PGD10-AT and significantly better than other defense methods. Although FS10-AT model is unable to beat PGD10-AT, it is significantly better than the standard and FGSM-AT models.
\par
We further evaluate the model robustness against PGD attacker under different attack iterations with a fixed $\epsilon = 0.002$, and the results are shown in Fig. \ref{fig3}. It is evident from the figure that both PGD10-AT and HAT10 maintain a fairly stable performance across the change in number of attack iterations from 10 to 100. We also observe that the proposed HAT10 consistently outperforms other baseline defenses for all attack iterations.

\vspace{-12pt}
\subsubsection{Ablation study}
\vspace{-5pt}

\begin{figure}[!t]
 \centering
  \includegraphics[width=0.5\textwidth, height = 3cm]{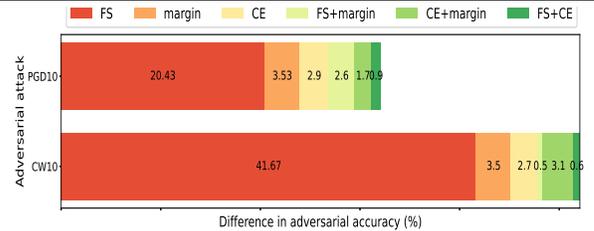}
  \vspace*{-0.62cm}
  \caption{Difference in accuracy (\%) between final proposed system using HAT and system using different components of the proposed loss combination for adversarial training.}
  \label{fig1}
  \vspace{-10pt}
\end{figure}

We perform an ablation study to examine the contribution of each loss function of our proposed defense technique. To do so, we employ each loss function individually and also their different possible combinations for adversary generation, and perform adversarial training to finally measure the adversarial accuracy of all those cases. Fig.~\ref{fig1} shows the difference in accuracy between our final proposed loss combination (CE+FS+margin) and each losses for both PGD10 and CW10 attacks. The effect of the losses on performance in increasing order is FS, margin, CE, FS+margin, CE+margin and FS+CE for PGD10 attack, and FS, margin, CE+margin, CE, FS+CE and FS+margin for CW10 attack, respectively. 

\begin{table}[]
\caption{Adversarial accuracy under transfer/black-box attacks.}
\vspace{-10pt}
\centering
\setlength{\tabcolsep}{7pt}
\begin{tabular}{c|cc|cc}
\Xhline{2.5\arrayrulewidth}
\multirow{2}{*}{Source} & \multicolumn{2}{c|}{PGD40-attack} & \multicolumn{2}{c}{CW40-attack} \\
\cline{2-3} \cline{4-5}  
                        & PGD10-AT        & HAT10       & PGD10-AT       & HAT10       \\ \Xhline{2.5\arrayrulewidth}
Standard                &     97.30         &       \textbf{97.58}      &      97.32       &    \textbf{97.54}         \\
FS10-AT                    &    96.47          &     \textbf{97.06}        &    96.74         &  \textbf{97.19} \\ \Xhline{2.5\arrayrulewidth}         
\end{tabular}\label{table2}
\vspace{-14pt}
\end{table}

\vspace{-12pt}
\subsection{Results on transfer/black-box attacks}
\vspace{-5pt}

We further evaluate the robustness of the HAT10 and PGD10-AT models under black-box setting, where the adversary has complete knowledge about the source model, but no or very limited knowledge about the target model (such as model architecture). The PGD40 and CW40 attacks are generated from the Standard and FS10-AT models and transferred to the PGD10-AT and HAT10 models. The corresponding results are summarized in Table \ref{table2}. We observe that for black-box attacks accuracy of our defense remains higher than PGD10-AT defense.

\vspace{-12pt}
\subsection{Sanity checks for gradient masking}
\vspace{-5pt}

To further verify if gradient masking (phenomenon where adversarially trained models provide false sense of robustness by learning to generate less useful gradients to adversarial attacks \cite{athalye2018obfuscated}) is occurring, we do sanity checks devised in \cite{athalye2018obfuscated}. We can say that our defense does not rely on gradient masking based on the following observations (a) The adversarial accuracy under black-box attacks (reported in Table \ref{table2}) remains higher than white-box attacks (reported in Table \ref{table1}) (b) In all our evaluations (Table \ref{table1}, Fig. \ref{fig2}, Fig. \ref{fig3}), our proposed defense yields lower adversarial accuracy for iterative PGD attacks as compared to one-step FGSM attack (c) Our defense accuracy approaches to zero (Fig. \ref{fig2}) if we increase the attack budget to a large value.

\vspace{-17pt}
\section{Conclusions}
\label{sec:foot}
\vspace{-7pt}

In this work, we proposed a hybrid adversarial training algorithm that provides diverse and stronger adversarial perturbations (by jointly utilizing multiple losses) for adversarial training to improve the robustness of deep speaker recognition system against various types of attacks.
We demonstrate the effectiveness of the proposed defense method through experiments on various white-box and black-box attacks with different perturbation budgets. In the future, we will extend this work for an end-to-end SV setting, and will evaluate under $l_1, l_2$ norm based attacks and also auditory masking based human imperceptible attacks.


%
%
%



\bibliographystyle{IEEEbib}
\bibliography{ref1}

\end{document}